\documentclass{svjour3}                     

\smartqed  

\usepackage{graphicx}
\usepackage{hyperref}
\usepackage{amsmath}
\usepackage{lmodern}
\usepackage{epstopdf}
\begin{document}

\title{Testing the CCC+TL Cosmology with Big Bang Nucleosynthesis
}


\author{Rajendra P. Gupta         \and
        Nikolaos Samaras 
}


\institute{Rajendra P. Gupta \at
              Department of Physics, University of Ottawa, Ottawa, Canada K1N 6N5 \\
              \email{rgupta4@uottawa.ca}           
           \and
           Nikolaos Samaras \at
              Department of Physics, University of Ottawa, Ottawa, Canada K1N 6N5\\
            \email{nsamaras@uottawa.ca}
}

\date{Received: date / Accepted: date}

\maketitle

\begin{abstract}
We investigate Big Bang nucleosynthesis (BBN) as a consistency test of the Covarying Coupling Constants plus Tired Light (CCC+TL; CTL) cosmology. In this framework, only quantities with explicit length dimensionality covary through a universal scaling function $f \left( z \right)$, while dimensionless constants and dimensionless ratios remain invariant. At the redshifts $z$ relevant to BBN, $f \left( z \right )$ approaches a constant plateau $f_{\text{max}} \left( z \right) \simeq 3$, and the tired-light contribution is negligible, so the early-time dynamics reduce to a global rescaling of dimensioned quantities. In particular, the Hubble expansion rate $H$ at fixed temperature $T$ satisfies $H_{\text{CTL}} \left( T \right) = f^{-1}_{\text{max}} H_{\Lambda\text{CDM}}\left( T\right)$, implying a longer interval $\Delta t$ between weak freeze-out and the onset of nucleosynthesis by the same factor. We find that BBN predictions are preserved provided the relevant interaction rates $\Gamma$ and decay rates scale with the same plateau factor as $H$, so that governing combinations such as $\Gamma\text{/}H$ and the neutron survival factor $\text{exp} \left( -\Delta t \text{/} {\tau}_n \right)$ remain invariant. Implementing these plateau rescalings in the Kawano/NUC123 network (via a single control parameter $\texttt{fctl} \equiv f_{\text{max}}$) yields identical light-element abundances for $\texttt{fctl}= 1$ ($\Lambda$CDM) and $\texttt{fctl} = 3\left( \text{CCC+TL} \right)$ that agree within $\sim 10^{-3} - 10^{-4}$ level, consistent with numerical precision. We also illustrate that adopting the lower late-time CCC+TL baryon density inferred from the Pantheon+ data fit can reduce the ${}^7\text{Li}$ discrepancy but simultaneously increases D/H, implying that BBN alone does not discriminate between the late-time baryon-density inferences considered here.
\end{abstract}

\section{Introduction}\label{intro}
\subsection{Big Bang Nucleosynthesis (BBN)}
BBN is the earliest empirically testable episode in cosmic history, linking microphysical reaction networks to the macroscopic expansion of the universe and yielding quantitative predictions for the primordial abundances of the light nuclides D, ${}^3$He, ${}^4$He, and ${}^7$Li. The intellectual roots of BBN trace back to the first attempts to connect the expanding, hot early universe with the formation of elements \cite{1946PhRv...70..572G}. In the late 1940s, Alpher and collaborators developed the “hot big bang” nucleosynthesis picture, emphasizing thermonuclear processing in a rapidly cooling radiation bath \cite{PhysRev.74.1577,1948PhRv...73..803A} and extending the theoretical framework in subsequent analyses of nuclear reaction flows in an expanding medium \cite{1948PhRv...74.1737A}. Although the early programme overreached in its aim to synthesize heavy elements, it established the central idea that the early universe would naturally produce substantial helium and trace light isotopes, while heavier elements would require stellar nucleosynthesis \cite{RevModPhys.29.547,1957PASP...69..201C}. 

BBN became a mature quantitative theory in the 1960s and 1970s, catalysed by the recognition that primordial helium provides a direct probe of the hot early fireball \cite{1966ApJ...146..542P} and by the first detailed network calculation  of light-element yields in standard Friedmann-Robertson-Walker cosmologies \cite{1967ApJ...148....3W}. These developments were quickly integrated into a broader cosmological framework in which the expansion rate $H(T)$, the baryon-to-photon ratio $\eta$, and weak-interaction freeze-out jointly determine the neutron-to-proton ratio and thus the ${}^4$He mass fraction $Y_p$, while deuterium and ${}^3$He track the competition between nuclear burning and the declining density during expansion \cite{1972gcpa.book.....W,1977PhLB...66..202S}. In this era, BBN also emerged as a sensitive diagnostic of new physics through its dependence on relativistic energy density (“effective number of neutrino species”), lepton asymmetry, and possible non-standard expansion histories \cite{1969SMFE....6..144S,1984ApJ...281..493Y,1996RPPh...59.1493S}. A defining feature of standard BBN (SBBN) is its predictive economy: given well-measured nuclear cross-sections  and standard weak rates, the primordial abundances depend primarily on $\eta$ (and modestly on the radiation content and neutron lifetime) \cite{1991ApJ...376...51W,2000PhR...333..389O,2001ApJ...552L...1B}. The reliability of these predictions  rests on continued progress in nuclear inputs and neutrino-decoupling physics, including refined thermonuclear reaction rate compilations  and sensitivity studies \cite{1999NuPhA.656....3A,DESCOUVEMONT2004203,2004PhRvD..70b3505C,2011RvMP...83..195A} and improved treatments of non-instantaneous neutrino decoupling and QED plasma effects (\cite{1982ApJ...252....1D,MANGANO2005221}). Parallel to these theoretical advances, community codes and benchmarks were developed to propagate nuclear and cosmological uncertainties into abundance predictions, from early implementations to widely used public calculations \cite{1992STIN...9225163K,1993ApJS...85..219S,2008CoPhC.178..956P,2012CoPhC.183.1822A}. 

Observationally, BBN is relevant because it anchors the cosmic baryon density and provides a stringent consistency check across epochs: $\eta$ inferred from primordial deuterium in metal-poor absorbers  \cite{1998ApJ...507..732B,2001ApJ...552..718O,2012MNRAS.425.2477P,2014APS..APRR11001C} can be directly compared to the baryon density independently inferred from cosmic microwave background (CMB) anisotropies \cite{2003ApJS..148....1B,2020A&A...641A...6P}. This BBN–CMB concordance has become a cornerstone of the standard cosmological model and a powerful lever arm for constraining extensions such as extra relativistic species or altered early-time expansion \cite{2007ARNPS..57..463S,2011ARNPS..61...47F,2016ApJ...830...55C,2018PhR...754....1P}. Primordial ${}^4$He, inferred from recombination-line spectroscopy of low-metallicity H II regions, provides complementary sensitivity to the expansion rate and lepton asymmetry \cite{2010ApJ...710L..67I,2015JCAP...07..011A,2014MNRAS.445..778I}. Together with deuterium, helium locks down the thermal history of the first minutes and quantifies the allowed room for beyond-standard-model effects. 

Despite these successes, BBN also highlights persistent tensions that motivate renewed theoretical scrutiny. Most notably, the long-standing “lithium problem”—the discrepancy between SBBN predictions  and the lower ${}^7\text{Li}$ abundances observed in metal-poor halo stars—has stimulated extensive work on stellar depletion, nuclear systematics, and new-physics remedies \cite{2011ARNPS..61...47F,1982A&A...115..357S,Ryan_2000,Asplund_2006,2010A&A...522A..26S}. Thus, BBN remains  simultaneously a triumph of early-Universe physics and an active testing ground: it connects particle interactions, nuclear astrophysics, and cosmology, and it supplies one of the most sensitive probes of any model that modifies reaction kinetics, the expansion rate, or the mapping between temperature and time in the pre-recombination Universe  \cite{2007ARNPS..57..463S,2016ApJ...830...55C,2018PhR...754....1P}.

In summary, BBN provides one of the earliest and most stringent probes of cosmology, linking microphysical processes at temperatures $T \sim 0.01$-$10$ MeV to present-day light-element abundances. In the standard $\Lambda$CDM framework, BBN predictions depend primarily on the baryon-to-photon ratio $\eta$, the Hubble expansion rate $H \left( T \right)$, and well-measured weak and nuclear reaction rates \cite{1992STIN...9225163K,1969ARA&A...7..553W,2017IJMPE..2641002C,2020JCAP...03..010F}.

\subsection{Covarying Coupling Constants (CCC)}
The covarying coupling constants framework traces its lineage to ideas in Dirac-style cosmology: Dirac’s proposal that the strengths of gravity and electromagnetism may evolve in a correlated way, together with the broader point—emphasized in later work—that allowing one dimensionful constant to vary generally implies that other dimensionful constants cannot consistently remain fixed. In CCC, this logic is taken seriously while keeping dimensionless constants (for example, the fine-structure constant) outside the scope of the principle.  

Historically, the conceptual thread begins  with Dirac \cite{1937Natur.139..323D} and continues through early developments by  Gilbert \cite{1956MNRAS.116..678G,1961Natur.192...57G} and by Cunato \& Londenquai \cite{1977ApJ...211..342C}, who explored cosmological consequences of evolving constants in generalized Dirac frameworks. In the decades since, a number of mathematically adjacent approaches have appeared. Of particular relevance are scale-invariant vacuum cosmologies developed by Bouvier (summarized in Maeder \cite{2023MNRAS.520.1447M,2017ApJ...847...65M,2019arXiv190210115M}, where effective time dependence in gravitational strength (and, in their formulation, the fine-structure constant as well) arises via Weyl-integrable rescalings of the metric. While the underlying motivations  differ from CCC, these theories likewise introduce non-trivial temporal evolution in gravitational coupling that can lead to phenomenology reminiscent of CCC-induced modification  to Friedmann evolution. Building on this general lineage, the CCC programme pursued here frames the co-evolution of dimensionful constants through a single governing function $f\left( t \right)$, showing how such correlated rescalings can yield effective contributions  that behave like dark matter and dark energy across cosmological and astrophysical settings.  

Time-variable $G$ scenarios have also been widely investigated in scalar–tensor gravity, most notably Jordan–Brans–Dicke theory \cite{1961PhRv..124..925B}. CCC overlaps with these models only at the broad level of permitting a dynamical gravitational coupling: structurally, CCC does not attribute the variations  of $G, c, h$ and related constants to a new scalar degree of  freedom with a canonical kinetic term. Instead, CCC proceeds phenomenologically, guided by dimensional consistency and local conservation principles, producing a distinct pattern of modification  to Einstein–Friedmann dynamics. In parallel, Weyl-based constructions—Weyl's original geometry and later Weyl-integrable formulations \cite{1918SPAW.......465W,1977ApJ...211..342C,2012CQGra..29o5015R}—also employ non-Riemannian structure to generate effective rescalings of physical units. CCC does not assume Weyl gauge symmetry or non-metricity, but the resulting effective behavior of the gravitational coupling can appear superficially similar in certain regimes. In this sense, CCC occupies a complementary niche: neither a Jordan–Brans–Dicke scalar–tensor model nor a Weyl-integrable geometric reformulation, but a correlated-variation framework in which dimensionful constants evolve together. An action-based perspective connecting $c\left( t \right)$ and $G \left( t \right)$ has been developed elsewhere \cite{2023Symm...15..709C}.  

Finally, it is worth noting that the literature placing empirical bounds on variations  of $G, c$ and other dimensionful parameters is extensive, spanning laboratory, Solar System, stellar, pulsar-timing, and cosmological probes. A key interpretive point for CCC, however, is that many such constraints are formulated within a “single-varying-constant” paradigm: one parameter is permitted to drift while others are implicitly held fixed. Citations for the $G$ variation studies include \cite{PhysRev.73.801,PhysRevLett.36.833,2014IJMPD..2342018S,1973Natur.241..519M,PhysRevD.41.1034,Corsico_2013,1996A&A...312..345D,10.1046/j.1365-8711.1999.02486.x} and 
\cite{1996PhRvL..77.1432T,2004PhRvL..93z1101WH,2015JCAP...10..029B,2017PTEP.2017d3E03O,2004PhRvL..92q1301C,2020EPJC...80..148A,2019ApJ...887L...1B,2018CQGra..35c5015H,2013MNRAS.432.3431P,2015CeMDA.123..325F,2018NatCo...9..289G,1988PhRvL..61.1151D,1994ApJ...428..713K,2019MNRAS.482.3249Z,2001PhRvD..65b3506G,2018PhRvD..97h3505W}. Among others, the potential variation of $c$ has been studied by \cite{1907AnP...328..197E,1957RvMP...29..363D,1993gr.qc....12017M,1993gr.qc.....6003M,1999PhRvD..59d3516A,1999PhRvD..59d3515B,1999PhLB..459..468A,2000PhRvD..62l3508A,2016EPJC...76..130M}. From the CCC viewpoint, these studies effectively set the common scaling function to unity by assumption, precluding the correlated evolution that CCC posits and rendering “one-constant-at-a-time” interpretations potentially misleading when translated into a covarying framework. 

\subsection{CCC+Tired Light (TL) phenomenology}
Tired light limitations  that led to the rejection of this concept, such as Compton scattering, time dilation, the Tolman brightness test, and the CMB isotropy, do not apply, as discussed in earlier papers e.g., \cite{2023Symm...15..709C,2024ApJ...964...55G,2024Univ...10..266G,2025Galax..13..108G,2025Galax..13..115G}, primarily because the tired light effect exists in parallel with the universe’s expansion. 

In the CCC framework, this tired-light component is not an ad hoc phenomenological term but a manifestation of an underlying vacuum microstructure that co-determines the fundamental dimensionful constants. The vacuum is treated as a medium with microscopic degrees of freedom, described by a coarse-grained order parameter $\Phi \left( t \right)$, through which photons propagate and lose an infinitesimal fraction of their energy in a non-scattering, non-dispersive way, while the universe also expands. The same microstructure governs  the effective speed of light $c \left( t \right)$ and gravitational coupling $G \left( t \right)$, so that they become functions  of $\Phi \left( t \right)$, whereas dimensionless constants such as the fine-structure constant $\alpha$ remain strictly invariant. 

A key structural ingredient of CCC is the invariant ratio $G \text{/}c^3 = \gamma$, motivated by the form of the Einstein–Hilbert action, which requires the prefactor $c^3 \text{/}G$ to be time-independent if the gravitational field equations  are to retain their standard form. This motivates a scaling ansatz in which only quantities with non-zero length dimensionality vary: if a quantity $X$ has length dimension $L^n$, then $X \left( t\right) = X \left( 0 \right) f^n \left( t\right)$ for some dimensionless microstructural scaling function $f \left( t \right)$. A simple realisation is $c \left( t \right) = c \left( 0 \right) f\left( t \right)$, $G \left( t \right) = G \left( 0 \right) f^3 \left( t \right)$, $\hbar \left( t\right) = \hbar \left( 0 \right) f^2 \left( t\right)$ and $k_B \left( t\right) = k_B\left( 0 \right) f^2\left( t\right)$, so that $G/c^3$ is constant, and Planck units acquire a natural microscopic interpretation. The length dimensionality rule is modified slightly for the expressions involving time differentiation or integration, such as for the Hubble rate with no length dimension scaling as $f^{-1} \left( t\right)$ as shown below. 

This microstructure admits a Kaluza-type geometric reading in which $\Phi \left( t \right)$ plays the role of an effective extra dimension whose geometry co-determines both light propagation and gravitational strength, echoing earlier unification attempts in higher-dimensional and induced-gravity theories \cite{1921SPAW.......966K,1967JETPL...5...24S,1997PhR...283..303O,2010grav.book.....P}. An important conceptual consequence is that observational tests that vary only a single constant while holding all others fixed are inconsistent with any such co-varying-constant framework: fixing one dimensional constant freezes the underlying degree of freedom and therefore forces all dependent constants to be constant by construction \cite{1994GReGr..26.1171D,2003AnHP....4S.347U,2017RPPh...80l6902M}. CCC therefore promotes a more coherent analysis in which the expansion rate, the tired-light contribution, and the co-variation of $c \left(t\right), G\left(t\right), \hbar\left(t\right)$ and $k_B\left(t\right)$ are treated consistently, while $\alpha$ and other dimensionless couplings remain unchanged  \cite{2023MNRAS.526.3987C}. 

BBN is one of the most essential tests to pass for any new cosmology model. The CCC+TL model modifies cosmology by allowing dimensional constants to covary with cosmic time while preserving dimensionless constants and ratios. It also replaces dark matter and dark energy with modified distance-redshift relations and a revised definition of the critical density. Given these departures, it is natural to ask whether CCC+TL alters the successful $\Lambda$CDM BBN predictions. 

The covarying coupling constant plus tired light model has already been successful in alleviating the ‘impossible early galaxy problem’ and fitting the SNe Ia Pantheon+ data \cite{2023AAS...24232604G}. Additionally, it is consistent with a) the BAO and CMB sound horizon observations  \cite{2024ApJ...964...55G}, b) galaxy formation time scales at cosmic dawn and time dilation \cite{2024Univ...10..266G}, c) galaxy rotation curves and galaxy cluster dynamics \cite{2025Galax..13..108G}, d) mass, size, density, and luminosity evolution of galaxies \cite{2025Galax..13..115G}, e) gravitational lensing and DESI findings of increasing dark energy density with redshift \cite{2026Symm...18..300G}, and cosmic chronometer compatibility (Gupta 2026b under review). 

This paper is structured as follows: Section \ref{sec:2} presents key features of the CCC+TL model relevant for BBN; in Section \ref{sec:3} we discuss the ingredients of BBN; Section \ref{sec:4} is devoted to the nuclear reaction rates and neutron lifetime; Section \ref{sec:5} details the testing of the new model with a modified Kawano/NUC123 code; Section \ref{sec:6} is used for discussion; and Section \ref{sec:7} provides the conclusion. 

\section{Key Features of the CCC+TL model relevant for BBN} \label{sec:2}
The CCC+TL model has been extensively discussed and applied to multiple cosmological and astrophysical observations  in several papers mentioned above. Thus, in this section, we will discuss the model features that are directly relevant to BBN. 

\subsection{Covarying coupling constants} 
Derived from local energy conservation laws applied to exploding stars \cite{2022MPLA...3750155G}, they can be considered as a generalization of Dirac’s large number hypothesis \cite{1937Natur.139..323D}) that predicted the evolution of the gravitational constant with cosmic time.  In CCC+TL, any quantity $X$ with net length dimensionality $n$ scales as
\begin{equation}
    X \left( z \right) = X \left( 0 \right) f \left( z \right)^n,
\end{equation}
where $f \left( z \right)$ is a universal scaling function. Examples include:
\begin{itemize}
    \item Speed of light $c \sim f \left( n = 1 \right)$
    \item Newton's constant $G \sim f^3 \left(n = 3 \right)$
    \item Planck's constant $\hbar \sim f^2 \left(n = 2 \right)$
    \item Boltzmann's constant $k_B \sim f^2 \left(n = 2 \right)$
\end{itemize}
Thus, dimensionless constants such as the fine-structure constant, mass, charge, gauge couplings, etc., are invariant in the CCC+TL cosmology. Ratios of dimensioned quantities are dimensionless, so they do not evolve. At high redshifts corresponding to BBN and recombination, $f \left( z \right)$ assumes its asymptotic constant value of approximately 3, depending on the observational data fit of the late universe or recombination. Thus, all dimensional scalings reduce to fixed multiplicative factors. Mass and charge do not evolve since they have no length dimension. The premise here is that as the universe expands at a macroscopic scale, it affects the measurement unit of length at the microscopic scale. 

\subsection{Expansion history} 
Despite its conceptual differences from the $\Lambda$CDM model, the CCC+TL expansion rate at high redshifts is proportional to the $\Lambda$CDM model, i.e., for the radiation-dominated Friedmann universe given by the Hubble parameter $H \left( T \right)$ at temperature $T$,
\begin{equation}
    H_{\text{CTL}} \left( T \right) \propto H_{\Lambda \text{CDM}} \left( T \right).
\end{equation}
Here, CCC+TL is abbreviated as CTL. It can be shown as follows:  
Friedmann equation and continuity equation in a flat universe for the two models are \cite{2023AAS...24232604G}: \\[17pt]
\underline{$\Lambda$CDM model}
\begin{align}
    H^2 &= \frac{8\pi G_0}{3c_0^2} \left(  \epsilon_{m,0} \left( 1\text{+} z \right)^3 \text{+} \epsilon_{r,0} \left( 1 \text{+} z\right)^4 \right) \text{+} \frac{\Lambda}{3}, \label{LCDM1}\\
    \dot{\epsilon} & \text{+} 3 \frac{\dot a}{a} \left(  \epsilon + P\right) = 0 \label{LCDM2}.
\end{align}
Here, the scale factor is  $a$, the current energy densities are $\epsilon_{m,0}$ for matter and $\epsilon_{r,0}$ for radiation, and pressure is $P$, with $a$ related to the observed redshift $z$ via $a = 1 \text{/} \left( 1 \text{+}z\right)$ and $\Lambda$ is the cosmological constant contribution to the energy density.\\[17pt]
\underline{CCC model}
\begin{align} 
    \left( H \text{+} \alpha \right)^2 &= \frac{8 \pi G_0}{3 c^2_0} \bigg(  \epsilon_{m,0} \left( 1 \text{+}z \right)^3  f \left(z \right)^{-1} \text{+} \epsilon_{r,0} \left( 1\text{+}z\right)^4 f\left(z \right)^{-2}\bigg),  \label{CCC1}\\[10pt]  
    \dot\epsilon &\text{+} \left( 3H + \alpha\right)\epsilon \text{+}3 \left( H \text{+} \alpha \right)P = 0 \label{CCC2}.
\end{align}
Here $\alpha$ is a constant defining the variation of the constants through $f \left( t \right) = \text{exp} \left( \alpha  \left( t-t_0\right) \right)$ with $t_0$ being the current time. Since $H$ increases rapidly with redshift, $\alpha$ can be neglected in Eq. \ref{CCC1} for high-redshift, early universe studies. The same is true about the continuity equation, Eq. \ref{CCC2}; it becomes the same as Eq. \ref{LCDM2}.

\begin{figure}
    \centering
    \includegraphics[width=\linewidth]{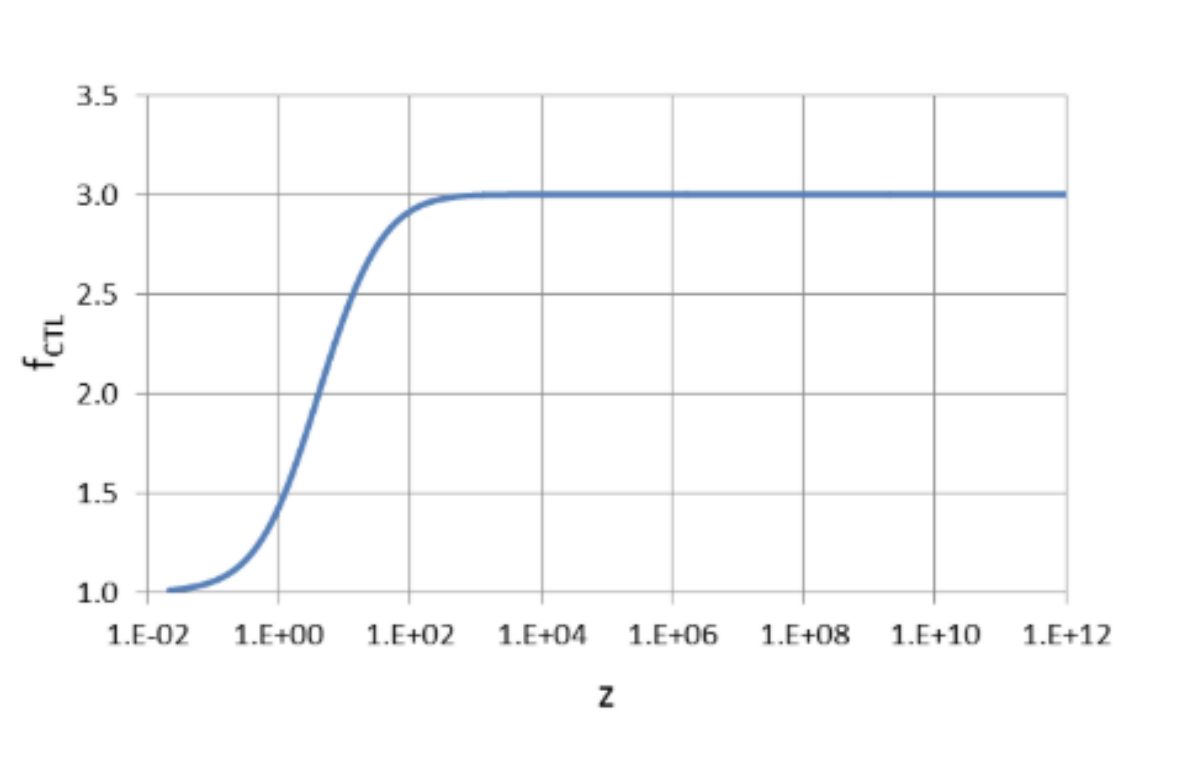}
    \caption{Variation of the function $f$ with $z$. It has a fixed value, $f_{\text{max}}$ at BBN redshifts.}
    \label{fctl_vs_z}
\end{figure}

Comparing Eq. \ref{LCDM1} and Eq. \ref{CCC1} in the radiation-dominated universe, we see $H_{\text{CCC}}= f \left( z\right)^{-1} H_{\Lambda\text{CDM}}$. Since $f\left( z \right)$ asymptotically approaches a maximum value $f_{\text{max}}$, and since cosmic temperature evolves as $\left( 1 \text{+} z\right)$, we get $H_{\text{CCC}} \left( T \right) = f^{-1}_{\text{max}}$.\\[17pt]
\underline{CCC+TL model}\\
The treatment of this model comprises expressions involving tired light \cite{2023AAS...24232604G}:
\begin{align}
    \int^{z_c}_0 \frac{dz}{\left( H_{c,0} \text{+} \alpha \right) \left( 1 \text{+} z\right)^{3\text{/}2} f\left( z\right)^{-\left( 1\text{/}2\right)}- \alpha}-[H_{t,0}]^{-1} \ln \bigg[ \frac{1\text{+}z}{1\text{+}z_c}\bigg] = 0.
\end{align}
Rewriting it, 
\begin{align}
    \left( 1\text{+}z \right) &\equiv \left( 1\text{+}z_c\right) \left( 1\text{+}z_t\right) \nonumber\\
    &=\left( 1\text{+}z_c\right) \cdot \text{exp} \bigg(  H_{t,0} \int^{z_c}_0 \frac{dz}{\left( H_{c,0} \text{+} \alpha \right) \left(1 \text{+}z \right)^{3\text{/}2}f \left( z \right)^{-\left( 1\text{/}2\right)}- \alpha}\bigg). \label{CCCTTL1}
\end{align}
\begin{figure}
    \centering
    \includegraphics[width=\linewidth]{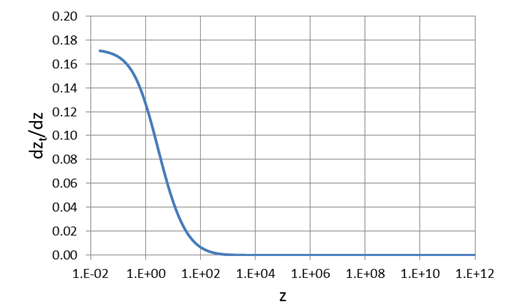}
    \caption{The tired light contribution to the observed redshift with typical $H_{c,0}$ and $\alpha$ parameters. It shows a complete absence of the tired-light effect during the BBN epoch.}
    \label{1+zt+vs+1_z}
\end{figure}
Here, $H_{c,0}$ is the Hubble constant corresponds to CCC and $H_{t,0}$ to TL with $H_0 = H_{c,0}\text{+}H_{t,0}$, with $z_{c}$ the CCC expanding Universe redshift and $z_t$ due to TL. $H_{t,0}$ is related to $H_{c,0}$ through
\begin{equation}
    H_{t,0} = \frac{\left( H_{c,0} \text{+}\alpha\right)}{2} \big( 3 \text{+} \frac{\alpha}{H_{c,0}}\big).
\end{equation}
The exp factor in Eq. \ref{CCCTTL1} represents the tired light redshift $\left(1 \text{+}z_t \right)$. Its behavior is shown in Fig. \ref{1+zt+vs+1_z}. Tired light has no effect at the BBN epoch. Therefore, we can use the same expressions for CCC+TL as for CCC; as in the case of CCC, $f\left( z \right)$ asymptotically approaches a maximum value $f_{\text{max}}$ - approximately 3, depending on the values of the parameters 
$H_{c,0}$ and $\alpha$  determined by fitting observational data such as Pantheon+ \cite{2018ApJ...859..101S,2022ApJ...938..110B}, Fig. \ref{fctl_vs_z}.

\begin{figure}
    \centering
    \includegraphics[width=\linewidth]{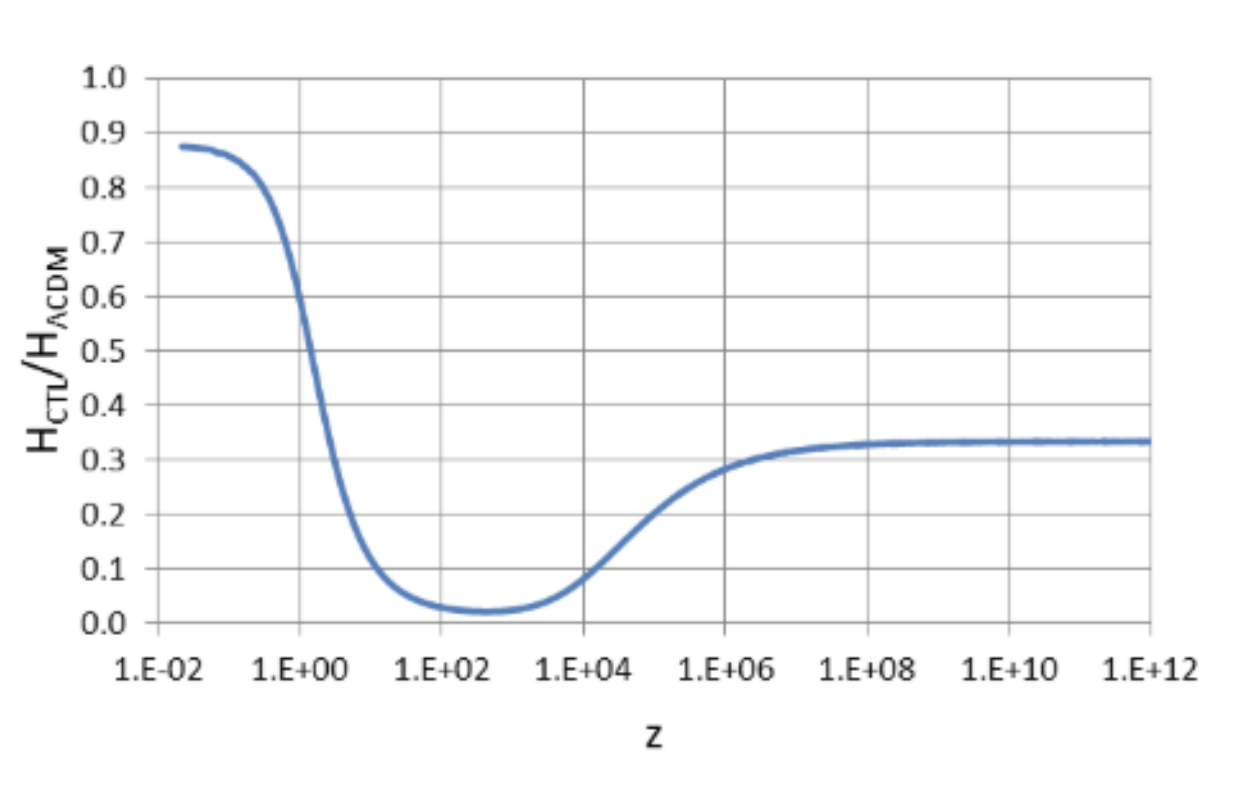}
    \caption{Variation of $H_{\text{CTL}}\text{/}H_{\Lambda \text{CDM}}$ with $z$. The ratio has a fixed value, 1/ $f_{\text{max}}$, at BBN redshifts.}
    \label{Hctl_HLCDM}
\end{figure}

As shown in an earlier paper \cite{2024ApJ...964...55G}, the temperature $T$ in the CCC+TL model evolves as $\left( 1 \text{+} z\right)$, i.e., the same as in the $\Lambda$CDM and CCC models. We can therefore conclude that $H_{\text{CTL}}\left( T\right) = f^{-1}_{\text{max}} H_{\Lambda\text{CDM}} \left( T\right)$ at the BBN epoch, Fig. \ref{Hctl_HLCDM}.

\begin{figure}
    \centering
    \includegraphics[width=\linewidth]{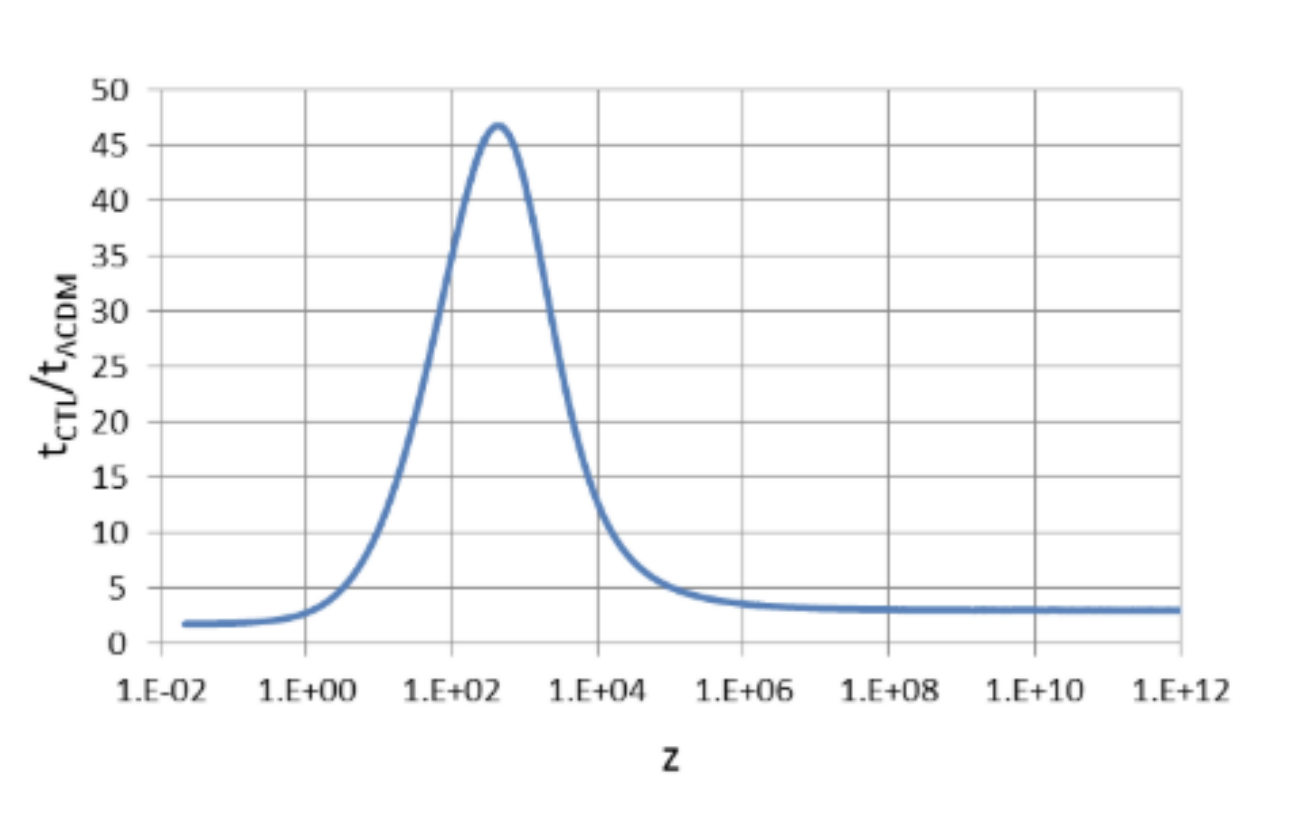}
    \caption{Variation of $t_{\text{CTL}}\text{/}t_{\Lambda \text{CDM}}$ with $z$. The ratio has a fixed value, $f_{\text{max}}$, at BBN redshifts.}
    \label{tctl_tLCDM}
\end{figure}

We also determined the evolution of cosmic time (age of the universe) with redshift. It is shown in Fig. \ref{tctl_tLCDM} as the ratio of $t_{\text{CTL}}\text{/}t_{\Lambda\text{CDM}}$, which approaches a constant value $f_{\text{max}}$.

\subsection{Critical density and baryon density today}
In a flat CCC+TL universe without dark matter or dark energy, the critical density is defined as 
\begin{equation}
    \epsilon^{\text{CTL}}_{c,0} = \frac{3c^2_0 \left( H_{c,0} \text{+}\alpha\right)^2}{8\pi G_0}= \frac{\left(H_{c,0} \text{+}\alpha \right)^2}{H^2_0}\epsilon^{\Lambda\text{CDM}}_{c,0}.
\end{equation}

Considering that from Pantheon+ data, $H_{c,0}$ = 59.5 km s\textsuperscript{-1} Mpc\textsuperscript{-1}, $\alpha$= $- 0.80$ $H_{c,0}$, and $H_0= 73.0$ km s\textsuperscript{-1} Mpc\textsuperscript{-1}, we get $\epsilon^{\text{CTL}}_{c,0} = 0.027 \epsilon^{\Lambda\text{CDM}}_{c,0}$. If we consider the same parameters at recombination, i.e., $H_{c,0} = 59.5$ km s\textsuperscript{-1} Mpc\textsuperscript{-1}, $\alpha = - 0.75 H_{c,0}$, and $H_0 =67.4$ km s\textsuperscript{-1} Mpc\textsuperscript{-1}, we get $\epsilon^{\text{CTL}}_{c,0} = 0.048 \epsilon^{\Lambda\text{CDM}}_{c,0}$. It means  that the critical energy density in the CCC+TL cosmology is in the range of the baryon energy density in the $\Lambda$CDM. Since the photon energy density is determined by the cosmic microwave background temperature of 2.7255 K, it yields the same photon energy density in both models. And, since number density, $n$, evolution is the same for baryons  and photons  when the total number of each is conserved, we get $\eta \equiv n_b \text{/}n_{\gamma}$ for the CCC+TL model, ranging from 56\% to 100\% of its $\Lambda$CDM value of $\cong$ 0.048 $\epsilon^{\Lambda\text{CDM}}_{c,0}$.

\section{Ingredients of BBN} \label{sec:3}
BBN predictions depend on the following dimensionless or effectively dimensionless quantities: 
\begin{enumerate}
    \item Baryon-to-photon ratio, $\eta = \frac{n_b}{n_{\gamma}}$. This is the dominant parameter controlling the abundances of deuterium and other light elements. As discussed above, it can differ in the CCC+TL model by a factor of 0.56 or more compared to $\Lambda$CDM. 
    \item Expansion-to-reaction-rate ratios, $\frac{\Gamma_{{n \leftrightarrow p}} \left( T \right)}{H \left( T \right)}$, where $\Gamma_{n \leftrightarrow p }$ denotes the weak interconversion rate between neutrons and protons.
    \item Energy ratios: The ratios, such as binding energies relative to thermal energy, e.g., $B_D \text{/} k_B T$ for deuterium formation, where $B_D$ is the deuterium binding energy, do not change between the two models as they are dimensionless; as discussed above, dimensionless quantities and dimensionless ratios of dimensioned quantities do not evolve in the CCC+TL cosmology. 
    \item Neutron lifetime: The neutron lifetime determines the fraction of neutrons after freeze-out (temperature too low for thermal equilibrium) that are able to form He and other light elements.
\end{enumerate}

BBN does not depend directly on present-day density parameters such as $\epsilon_b$, except insofar as they map onto $\eta$.

\section{Weak interaction rates and neutron lifetime} \label{sec:4}
\subsection{\textit{Neutron–proton interconversion rate}}

It can be written schematically as \cite{2000PhR...333..389O,1989RvMP...61...25B,1990eaun.book.....K} 

\begin{equation}
    \Gamma_{n \leftrightarrow p} \left( T \right) \propto G_F \left( t \right)^2 T^5 \times F \left( \frac{\Delta m_{np}c^2}{k_BT} \right),
\end{equation}
where $G_F \left( t\right)$ is the Fermi constant (potentially time-dependent in a general theory), $\Delta m_{np}$ is the neutron–proton mass difference, $F$ is a dimensionless phase-space function. In a pedagogical approach, we may write it as neutrinos and antineutrinos mediating the back-and-forth conversion of protons  and neutrons  via the weak nuclear force: $\Gamma_{n \leftrightarrow p} = n_{\nu}c \sigma_w$. Here $n_{\nu}$ is the neutrino number density, and $\sigma_w$ is the weak interaction cross-section. Multiplying and dividing the right-hand side by the neutrinos’ energy $E_{\nu}$, we may write $\Gamma_{n \leftrightarrow p} =\epsilon_{\nu}c \sigma_w \text{/}E_{\nu}$. Considering that neutrinos are relativistic particles, their energy density $\epsilon_{\nu} \sim f^{-2}$ (see Eq. \ref{CCC1}) and energy $E_{\nu} \sim f^2$. With $c\sim f$ and $\sigma \sim f^2$, we can write $\Gamma_{n \leftrightarrow p} \sim f^{-1}_{\text{max}}$. Thus, 
\begin{equation}
    \biggr[ \frac{\Gamma_{n \leftrightarrow p}\left( T\right)}{H \left( T \right)} \biggr]_{\text{CTL}} = \biggr[ \frac{f^{-1}_{\text{max}}\Gamma_{n \leftrightarrow p} \left( T \right)}{f^{-1}_{\text{max}}H\left( T \right)} \biggr]_{\Lambda \text{CDM}} = \biggr[ \frac{\Gamma_{n \leftrightarrow p} \left( T \right)}{H \left(T \right)}\biggr]_{\Lambda\text{CDM}}.
\end{equation}
We conclude that BBN is unaffected due to the expansion-to-reaction rate ratio. In other words, the reaction rates follow the Hubble expansion rate. We may generalize it as an ansatz to all rates, such as interaction and decay rates, that they have a scaling symmetry consistent with the Hubble expansion rate. 

\begin{figure}
    \centering
    \includegraphics[width=0.8\linewidth]{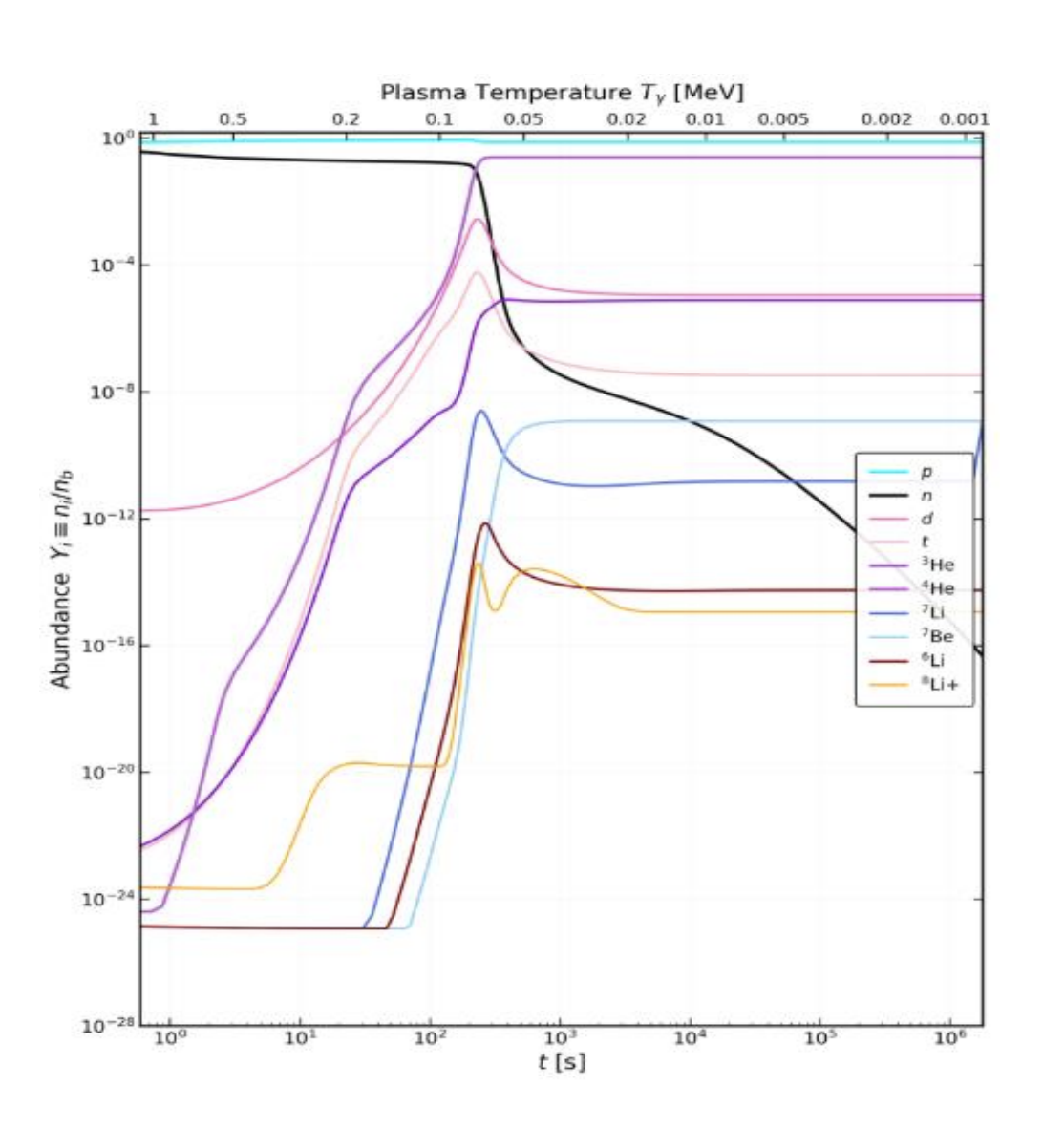}
    \caption{Abundance of light elements’ nucleosynthesis with time using the $\Lambda$CDM clock. The abundance cut-off was set at $10^{-25}$.  The temperature scales are the same in the $\Lambda$CDM and CCC+TL models, but the clock is three times slower in CCC+TL.}
    \label{Yi}
\end{figure}

\subsection{\textit{Deuterium bottleneck – Neutron survival}}
Formation of deuterium is the crucial step in the synthesis of elements from protons and neutrons. An excessive number of high-energy photons  at high temperatures quickly photodissociate newly formed deuterium, which is essential for the formation of helium and other elements. However, the Hubble expansion cools the universe, and at about 0.1 MeV ($T_{\text{BBN}}$), the energetic photon numbers are reduced enough to allow the formation of such elements.  But the lifetime of neutrons of about 880 seconds ($\tau_n$) leads to not all neutrons at freeze-out being available to form elements. The cooling time $\Delta t_{\text{cool}}$ from freeze-out to $T_{\text{BBN}}$ is thus crucial in determining the fraction of neutrons  forming the element through the proportionality factor exp$\left( - \Delta t_{\text{cool}}\text{/}\tau_n \right)$.

The temperature in the expanding universe in the CCC+TL and $\Lambda$CDM models is given by: 
\begin{align}
     T &= T_{\text{CMB}} \left( 1 \text{+}z \right) = T_{\text{CMB}}\text{/}a, \text{ i.e.,}\nonumber\\[10pt] 
     \frac{d T}{dt} &= -\frac{\dot a}{a^2}T_{\text{CMB}} = -HT_{\text{CMB}} \left( 1\text{+}z\right).
\end{align}
Since, as shown above, $H_{\text{CCC}}\left( T\right) = f^{-1}_{\text{max}} H_{\Lambda \text{CDM}} \left( T \right)$, for the same $dT$, $dt_{\text{CTL}} = f_{\text{max}} {dt}_{\Lambda\text{CDM}}$. Thus, a reduced Hubble expansion rate by a factor $f_{\text{max}}$ means  a longer cooling time by the same factor and a concomitant reduction in Helium formation. 

We have assumed here that the neutron decay rate, governed by the neutron lifetime, is unaffected by covarying coupling constants or changes in the Hubble expansion rate. The neutron lifetime expression is \cite{2014arXiv1411.3687W}: 
\begin{equation}
    \tau_n = \bigg( \frac{2 \pi^3 \hbar^7}{m_e^5 c^4 f_R} \bigg) \frac{1}{G^2_V \text{+} 3G^2_A}.
\end{equation}

\begin{figure}
    \centering
    \includegraphics[width=0.7\linewidth]{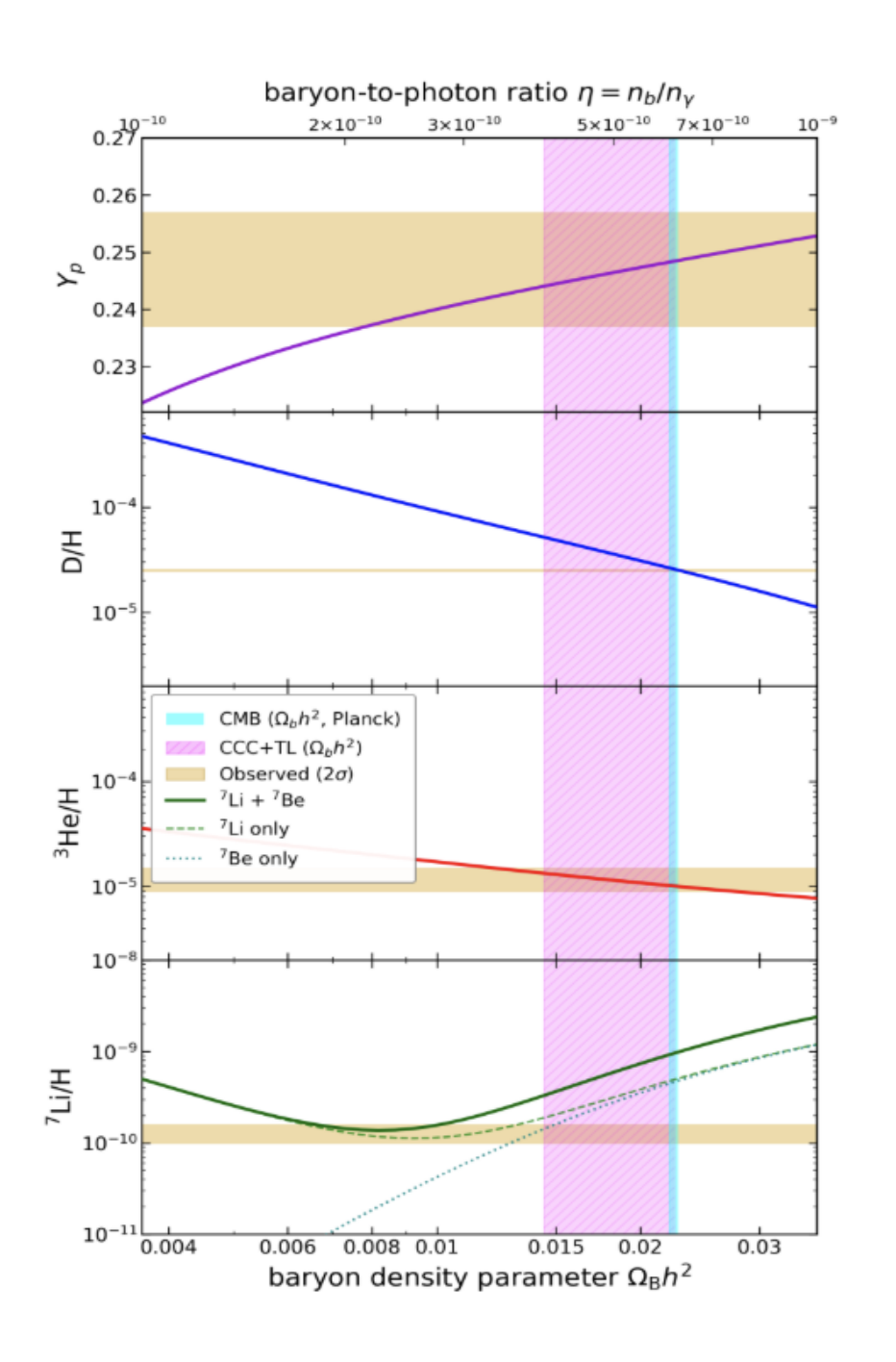}
    \caption{Schramm plot of BBN elemental abundances. The CCC+TL vertical column shows the spread in the values of baryon density calculated using Pantheon+ (lower value edge) and the CMB sound horizon angular size fit (higher value edge). The higher value edge overlaps with the Planck value.}
    \label{omegab}
\end{figure}
Here $f_R$ is a phase space factor that includes final state and radiative corrections, and $m_e$ is the electron mass. The nucleon vector and axial vector effective weak coupling constants $G_V$ and $G_A$ determine the neutron decay rate and therefore the neutron lifetime. The scaling relation for $\tau_n$ is not easy to determine in the CCC+TL universe. Nevertheless, using the rate symmetry ansatz mentioned above, the neutron decay rate scales the same as the Hubble rate, i.e., as 1/$f_{\text{max}}$, and therefore $\tau_n \sim f_{\text{max}}$. Alternatively, we may consider the neutron decay rate to be mediated by relativistic particles and thus have the same scaling as $H$, as we have discussed above. Thus, the proportionality factor exp$\left( -\Delta t_{\text{cool}}/\tau_n\right)$ that determines the fraction of neutrons  forming elements remains  unchanged in CCC+TL cosmology with respect to $\Lambda$CDM; the numerator and denominator in the argument of the exponential functions are both multiplied by $f_{\text{max}}$, and therefore cancel. 

\section{Testing with Kawano/NUC123} \label{sec:5}
We decided to test CCC+TL cosmology with a simple BBN code, Kawano/NUC123 \cite{1967ApJ...148....3W}.  It is a well-proven BBN code with limited nuclear reaction data, but it is transparent enough for the modifications  required for our purpose.  The idea is not to achieve high precision of AlterBBN \cite{2012CoPhC.183.1822A}, but to test whether the modifications  we have discussed above are meaningful and implementable. The changes made to the NUC123 code (FORTRAN 77/90) are as follows, where $\texttt{fCTL} \equiv f_{\text{max}}$: \vspace{10pt} \\ 
\texttt{SUBROUTINE rate4(t9):} scale reaction rates
\begin{align*}
    \texttt{f}  \texttt{(65:} &\texttt{88)}\\
    &\texttt{f} \texttt{(65:88)=f(65:88) / fctl}
\end{align*}
\textbf{Module} \texttt{driver.f}\vspace{5pt}\\
\texttt{SUBROUTINE start:} scale initial time \texttt{t}, timestep \texttt{dt}, and neutron lifetime \texttt{tau}
\begin{align*}
    \texttt{t} &= \texttt{fctl/(const1*uni\%t9)**2}\\
    \texttt{dt} &= \texttt{fctl*dt1}\\     
    \texttt{tau} &= \texttt{fctl*tau}
\end{align*}
\texttt{SUBROUTINE derivs:} scale the expansion rate
\begin{align*}
    \texttt{hubc} &\texttt{st} \\
    & \texttt{     hubcst = hubcst/fctl}
\end{align*}
\texttt{SUBROUTINE rate0:} divide the (constants) radioactive decay rates by \texttt{fctl}
\begin{align*}
    \texttt{DO } & \texttt{i=2,11} \\
    &\texttt{f} \texttt{(i) / fctl} \\
    \texttt{END} & \texttt{ DO}
\end{align*}
\texttt{SUBROUTINE rate1:} no addition of any \texttt{/fctl} in \texttt{f(1)} as it is already included through its dependence on \texttt{tau}.\\
\textbf{Module} \texttt{reactions.f90}\\
\texttt{SUBROUTINE rate2(t9):} scale reaction rates 
\begin{align*}
    \texttt{f} \texttt{(12:} &\texttt{34)}\\
    &\texttt{f} \texttt{(12:34)=f(12:34) / fctl}
\end{align*}
\texttt{SUBROUTINE rate3(t9):} scale reaction rates 
\begin{align*}
    \texttt{f(35:} & \texttt{64)}\\
     &\texttt{f(35:64)=f(35:64) / fctl }
\end{align*}
Reverse reaction rates \texttt{r} are computed elsewhere from forward reaction rates \texttt{f} by detailed balance. Thus, they inherit the same scaling automatically.\\
\textbf{Module} \texttt{variables.f90:} Declare \texttt{fctl}
\par \texttt{REAL, SAVE :: fctl}\\
\textbf{Module} \texttt{bbn.f90:} Read \texttt{fctl} and \texttt{eta} from a \texttt{.dat} file. 
The above modifications  are all that is needed to test the BBN compliance of the CCC+TL model using Kawano/NUC123 code. The results we obtain with \texttt{fctl=1} (no change, i.e., the $\Lambda$CDM model) and \texttt{fctl=3} (the CCC+TL model) are the same except for a 3\textsuperscript{rd}rd or 4\textsuperscript{th} significant figure difference in the elemental abundances, most likely due to numerical rounding errors. Results are presented in Fig. \ref{Yi} and Fig. \ref{omegab}.  

\section{Discussion} \label{discussion} \label{sec:6}
A central requirement for any alternative cosmology is that it preserves the key empirical successes of standard Big Bang nucleosynthesis (SBBN), which links microphysical weak and nuclear reaction kinetics to the macroscopic expansion history and predicts the primordial abundances of D, ${}^3$He, ${}^4$He, and ${}^7$Li with a small set of inputs \cite{1967ApJ...148....3W,1991ApJ...376...51W,2000PhR...333..389O,2016ApJ...830...55C,2018PhR...754....1P}. In the standard BBN, the dominant control parameter is the baryon-to-photon ratio $\eta$, with subleading dependence on the expansion rate (often framed via $N_{\text{eff}}$) and the neutron lifetime \cite{2007ARNPS..57..463S,2011ARNPS..61...47F,2017IJMPE..2641002C,2020JCAP...03..010F}. The purpose of this work has been to assess whether the CCC+TL framework—where only quantities with explicit length dimensionality covary through a single scaling function $f\left( z \right)$ while dimensionless constants and ratios remain invariant—remains consistent with these well-tested BBN predictions. 

The key simplifying feature for BBN in CCC+TL is that the relevant high-redshift epoch lies on an asymptotic plateau where $f \left( z \right) \rightarrow f_{\text{max}} \approx$ constant, so the model reduces to a global rescaling symmetry for dimensioned quantities. In this regime, the CCC+TL expansion rate satisfies $H_{\text{CTL}} \left( T\right) =f^{-1}_{\text{max}} H_{\Lambda\text{CDM}}\left( T \right)$, while the temperature–redshift relation retains  the standard scaling $T \propto  \left(1 \text{+} z \right)$, preserving the thermodynamic milestones (freeze-out and the deuterium bottleneck) at the same temperatures as in $\Lambda$CDM. 

BBN then reduces to the behavior of \textit{dimensionless} (or effectively dimensionless) governing combinations. First, binding-energy thresholds enter as ratios such as $B_D$/$k_B T$, which remain unchanged because the CCC+TL model keeps dimensionless ratios invariant by construction. Second, weak freeze-out and related kinetics depend primarily on ratios like $\Gamma_{n \leftrightarrow p} \left( T \right)$/$H\left(T\right)$. The scaling argument shows that, on the plateau, reaction/interaction rates inherit the same $f^{-1}_{\text{max}}$ factor as $H$, leaving $\Gamma$/$H$ unchanged and therefore preserving the neutron-to-proton ratio at freeze-out. Third, although a smaller $H$ lengthens  the cooling interval $\Delta t$, this does not change the neutron survival fraction if decay rates (including neutron decay) scale in the same way. In that case the neutron lifetime scales as $t_n \propto f_{\text{max}}$, so the factor exp$\left(\Delta t \text{/} t_n \right)$ remains  unchanged. We treat this lifetime scaling as an ansatz, motivated by a broader “rate symmetry” principle and by the fact that BBN is governed mainly by ratios rather than absolute times. 

These theoretical expectations  are supported by the practical test performed here: implementing the CCC+TL plateau rescalings in Kawano/NUC123 by scaling $H$, time steps and the (assumed) relevant decay/reaction rates by the appropriate powers of $f_{\text{max}}$, we find that the predicted elemental abundances for $f_{\text{max}} = 3$ are indistinguishable from the $\Lambda$CDM case $f_{\text{max}} = 1$ up to $\sim$ 3rd-4th significant figure differences consistent with numerical rounding. 

Finally, Fig. \ref{omegab} highlights that interpreting BBN constraints in CCC+TL requires care in mapping late-time density parameters to $\eta$. Because BBN constraints act primarily on $\eta$ (not directly on present-day $\Omega_b$ definitions), the model remains  consistent provided baryon–photon number conservation holds and $\eta$ is fixed to the recombination value inferred from CMB anisotropies \cite{2003ApJS..148....1B,2020A&A...641A...6P}. This also clarifies why adopting the lower CCC+TL baryon-density estimate may reduce the ${}^7$Li discrepancy while simultaneously worsening deuterium, implying that BBN alone does not uniquely select between the late-time baryon-density inferences considered here. 

BBN is preserved in the CCC+TL cosmology provided decay and interaction rates obey the same plateau scaling as the Hubble rate $H$, not necessarily based on our pedagogical reasoning. It may be considered an ansatz or prediction.

\section{Conclusion} \label{conclusion} \label{sec:7}
We conclude that the CCC+TL model is consistent with the BBN-predicted primordial helium and other light-element observations. We show that parameters that determine the abundances of such elements are the same in the CCC+TL and $\Lambda$CDM models:
\begin{enumerate}
    \item Energy ratios and other ratios are dimensionless and therefore are unaffected by covarying coupling constants.
    \item Thermodynamics is unchanged; freeze-out and nucleosynthesis temperatures are unaltered.
    \item All the rates have scaling symmetry with the Hubble expansion rates, leaving the BBN equations  unchanged as they are defined directly or indirectly with respect to the Hubble rate. 
    \item In comparison to the $\Lambda$CDM values, the CCC+TL low-end baryon density reduces the ${}^7\text{Li}\text{/}$H by a factor of $\approx$ 2.6, i.e., reducing the lithium discrepancy. At the same time, it increases the $D\text{/}$H by a factor of 2, thus creating the deuterium discrepancy. The CCC+TL high-end baryon density is about the same as the $\Lambda$CDM value, yielding the same abundances for the two models. 
    \item  We infer that BBN does not constrain either of the two CCC+TL baryon density values and therefore does not determine preference for either model. 
\end{enumerate}

\bibliography{biblio}

\begin{thebibliography}{99}

\bibitem{1946PhRv...70..572G}Gamow, G. \emph{Expanding Universe and the Origin of Elements}, \href{https://ui.adsabs.harvard.edu/abs/1946PhRv...70..572G} {Physical Review \textbf{70} (1946) pg. 572-573}

\bibitem{PhysRev.74.1577}Alpher, R. A. \emph{A Neutron-Capture Theory of the Formation and Relative Abundance of the Elements.} \href{https://link.aps.org/doi/10.1103/PhysRev.74.1577}{Physical Review \textbf{74} (1948) pg. 1577-1589}

\bibitem{1948PhRv...73..803A}Alpher, R. A., Bethe, H., and Gamow, G. \emph{The Origin of Chemical Elements.} \href{https://ui.adsabs.harvard.edu/abs/1948PhRv...73..803A}{Physical Review \textbf{73} (1948) pg. 803-804}

\bibitem{1948PhRv...74.1737A}Alpher, Ralph A., and Herman, Robert C. \emph{On the Relative Abundance of the Elements.} \href{https://ui.adsabs.harvard.edu/abs/1948PhRv...74.1737A}{Physical Review \textbf{74} (1948) pg. 1737-1742}
\bibitem{RevModPhys.29.547}Burbidge, E. Margaret, Burbidge, G. R., Fowler, William A., and Hoyle, F. \emph{Synthesis of the Elements in Stars.} \href{https://link.aps.org/doi/10.1103/RevModPhys.29.547}{Rev. Mod. Phys. \textbf{29} (1957) pg. 547-650}

\bibitem{1957PASP...69..201C}Cameron, A.G.W. \emph{Nuclear Reactions in Stars and Nucleogenesis} \href{https://ui.adsabs.harvard.edu/abs/1957PASP...69..201C}{Nuclear Reactions in Stars and Nucleogenesis \textbf{} (1957) pg. 201}

\bibitem{1966ApJ...146..542P}Peebles, P.J.E. \emph{Primordial Helium Abundance and the Primordial Fireball. II} \href{https://ui.adsabs.harvard.edu/abs/1966ApJ...146..542P}{ApJ \textbf{146} (1966) pg. 542}

\bibitem{1967ApJ...148....3W}Wagoner, Robert V., Fowler, William A., and Hoyle, F.\emph{On the Synthesis of Elements at Very High Temperatures,} \href{https://ui.adsabs.harvard.edu/abs/1967ApJ...148....3W}{ApJ \textbf{148} (1967) pg. 3}

\bibitem{1972gcpa.book.....W}Weinberg, Steven \emph{Gravitation and Cosmology: Principles and Applications of the General Theory of Relativity} \href{https://ui.adsabs.harvard.edu/abs/1972gcpa.book.....W}{ \textbf{} (1972)}

\bibitem{1977PhLB...66..202S}Steigman, Gary, and Schramm, David N., and Gunn, James E.\emph{Cosmological limits to the number of massive leptons.} \href{https://ui.adsabs.harvard.edu/abs/1977PhLB...66..202S}{Physics Letters B \textbf{66} (1977) pg. 202-204}

\bibitem{1969SMFE....6..144S}Shvartsman, D. \emph{Reviews and bibliography.} \href{https://ui.adsabs.harvard.edu/abs/1969SMFE....6..144S}{Soil Mechanics and Foundation Engineering \textbf{6} (1969) pg. 144-145}

\bibitem{1984ApJ...281..493Y}Yang, J., Turner, M.S., Steigman, G., Schramm, D.N., and Olive, K.A. \emph{Primordial nucleosynthesis: a critical comparison of theory and observation.} \href{https://ui.adsabs.harvard.edu/abs/1984ApJ...281..493Y}{ApJ \textbf{281} (1984) pg. 492-511}

\bibitem{1996RPPh...59.1493S}Sarkar, Subir\emph{Big bang nucleosynthesis and physics beyond the standard model.} \href{https://ui.adsabs.harvard.edu/abs/1996RPPh...59.1493S}{Reports on Progress in Physics \textbf{59} (1996) pg. 1493-1609}

\bibitem{1991ApJ...376...51W}Walker, Terry P., Steigman, Gary, Schramm, David N., Olive, Keith A., and Kang, Ho-Shik. \emph{Primordial Nucleosynthesis Redux} \href{https://ui.adsabs.harvard.edu/abs/1991ApJ...376...51W}{ApJ \textbf{376} (1991) pg. 51}

\bibitem{2000PhR...333..389O}Olive, K.A., {Steigman}, G., and {Walker}, T.P. \emph{Primordial nucleosynthesis: theory and observations.} \href{https://ui.adsabs.harvard.edu/abs/2000PhR...333..389O}{Physics Reports \textbf{333} (2000) pg. 389-407}

\bibitem{2001ApJ...552L...1B}Burles, Scott, Nollett, Kenneth M., and Turner, Michael S. \emph{Big Bang Nucleosynthesis Predictions for Precision Cosmology} \href{https://ui.adsabs.harvard.edu/abs/2001ApJ...552L...1B}{ApJ Letters \textbf{552} (2001) pg. L1-L5}

\bibitem{1999NuPhA.656....3A} Angulo, C. et al., \emph{A compilation of charged-particle induced thermonuclear reaction rates.} \href{https://ui.adsabs.harvard.edu/abs/1999NuPhA.656....3A/abstract}{Nucl. Phys. A \textbf{656} (1999) pg. 3-183}

\bibitem{DESCOUVEMONT2004203}Descouvemont, P., Adahchour, A., Angulo, C., Coc, A., and Vangioni-Flam, E. \emph{Compilation and R-matrix analysis of Big Bang nuclear reaction rates.} \href{https://www.sciencedirect.com/science/article/pii/S0092640X04000282}{Atomic Data and Nuclear Data Tables \textbf{88} (2004) pg. 203-236}

\bibitem{2004PhRvD..70b3505C}Cyburt, R. H. \emph{Primordial nucleosynthesis for the new cosmology: Determining uncertainties and examining concordance.} \href{https://ui.adsabs.harvard.edu/abs/2004PhRvD..70b3505C}{Physics Review D. \textbf{70} (2004) pg. 2}

\bibitem{2011RvMP...83..195A}Adelberger, E.G., et al. \emph{Solar fusion cross sections. II. The pp chain and CNO cycles} \href{https://ui.adsabs.harvard.edu/abs/2011RvMP...83..195A}{Reviews of Modern Physics \textbf{83} (2011) pg. 195-196}

\bibitem{1982ApJ...252....1D}Dicus, D.A., Letaw, J.R., Teplitz, D.C., and Teplitz, V.L. \emph{Effects of proton decay on the cosmological future} \href{https://ui.adsabs.harvard.edu/abs/1982ApJ...252....1D}{\textbf{252} (1982) pg. 1-9}

\bibitem{MANGANO2005221}Mangano, G., Miele, G., Pastor, S., Pinto, T., Pisanti, O., and Serpico, P. D. \emph{Relic neutrino decoupling including flavour oscillations.} \href{https://www.sciencedirect.com/science/article/pii/S0550321305008291}{Nuclear Physics B \textbf{729} (2005) pg. 221-234}

\bibitem{1992STIN...9225163K}Kawano, Lawrence. \emph{Let's go: Early universe 2. Primordial nucleosynthesis the computer way} \href{https://ui.adsabs.harvard.edu/abs/1992STIN...9225163K}{(1992)}

\bibitem{1993ApJS...85..219S}Smith, Michael S., Kawano, Lawrence H., and Malaney, Robert A. \emph{Experimental, Computational, and Observational Analysis of Primordial Nucleosynthesis.} \href{https://ui.adsabs.harvard.edu/abs/1993ApJS...85..219S}{ApJS \textbf{85} (1993) pg. 219}

\bibitem{2008CoPhC.178..956P}Pisanti, O., Cirillo, A., Esposito, S., Iocco, F., Mangano, G., Miele, G., and Serpico, P.D. \emph{PArthENoPE: Public algorithm evaluating the nucleosynthesis of primordial elements.} \href{https://ui.adsabs.harvard.edu/abs/2008CoPhC.178..956P}{Computer Physics Communications\textbf{178} (2008) pg. 12}

\bibitem{2012CoPhC.183.1822A}Arbey, A. \emph{AlterBBN: A program for calculating the BBN abundances of the elements in alternative cosmologies.} \href{https://ui.adsabs.harvard.edu/abs/2012CoPhC.183.1822A}{Computer Physics Communications \textbf{183} (2012) pg. 1822-1831}

\bibitem{1998ApJ...507..732B}Burles, Scott, and Tytler, David \emph{he Deuterium Abundance toward QSO 1009+2956.} \href{https://ui.adsabs.harvard.edu/abs/1998ApJ...507..732B}{ApJ \textbf{507} (1998) pg. 732-744}

\bibitem{2001ApJ...552..718O}O'Meara, John M., Tytler, David, Kirkman, David, Suzuki, Nao, Prochaska, Jason X., Lubin, Dan, and Wolfe, Arthur M. \emph{The Deuterium to Hydrogen Abundance Ratio toward a Fourth QSO: HS 0105+1619} \href{https://ui.adsabs.harvard.edu/abs/2001ApJ...552..718O}{ApJ \textbf{552} (2001) pg. 718-730}


\bibitem{2012MNRAS.425.2477P}Pettini, Max, and Cooke, Ryan. \emph{A new, precise measurement of the primordial abundance of deuterium.} \href{https://ui.adsabs.harvard.edu/abs/2012MNRAS.425.2477P}{MNRAS \textbf{425} (2012) pg. 2477-2486}

\bibitem{2014APS..APRR11001C}Cooke, Ryan. \emph{Primordial deuterium at the per cent level.} \href{https://ui.adsabs.harvard.edu/abs/2014APS..APRR11001C}{APS April Meeting Abstracts \textbf{2014} (2014) pg. R11.001}

\bibitem{2003ApJS..148....1B}Bennett, C.L., et al. \emph{First-Year Wilkinson Microwave Anisotropy Probe (WMAP) Observations: Preliminary Maps and Basic Results.} \href{https://ui.adsabs.harvard.edu/abs/2003ApJS..148....1B}{ApJS \textbf{148} (2003) pg. 1-27}

\bibitem{2020A&A...641A...6P}Planck Collaboration \emph{Planck 2018 results. VI. Cosmological parameters.} \href{https://ui.adsabs.harvard.edu/abs/2020A&A...641A...6}{Astronomy and Astrophysics \textbf{641} (2020) pg. A6}

\bibitem{2007ARNPS..57..463S}Steigman, Gary. \emph{Primordial Nucleosynthesis in the Precision Cosmology Era.} \href{https://ui.adsabs.harvard.edu/abs/2007ARNPS..57..463S}{Annual Review of Nuclear and Particle Science \textbf{57} (2007) pg. 463-491}

\bibitem{2011ARNPS..61...47F}Fields, Brian D. \emph{The Primordial Lithium Problem.} \href{https://ui.adsabs.harvard.edu/abs/2011ARNPS..61...47F}{Annual Review of Nuclear and Particle Science \textbf{61} (2011) pg. 47-68}

\bibitem{2016ApJ...830...55C} Cyburt, R. H.et al. \emph{Dependence of X-Ray Burst Models on Nuclear Reaction Rates} \href{https://ui.adsabs.harvard.edu/abs/2016ApJ...830...55C}{ApJ \textbf{830} (2016) pg. 55}

\bibitem{2018PhR...754....1P}Pitrou, Cyril, Coc, Alain, Uzan, Jean-Philippe, and Vangioni, Elisabeth. \emph{Precision big bang nucleosynthesis with improved Helium-4 predictions.} \href{https://ui.adsabs.harvard.edu/abs/2018PhR...754....1P}{Physical Reports \textbf{754} (2018) pg. 1-66 }

\bibitem{2010ApJ...710L..67I}Izotov, Yuri I., and Thuan, Trinh X. \emph{The Primordial Abundance of $^{4}$He: Evidence for Non-Standard Big Bang Nucleosynthesis.} \href{https://ui.adsabs.harvard.edu/abs/2010ApJ...710L..67I}{ApJL \textbf{710} (2010) pg. L67-L71}

\bibitem{2015JCAP...07..011A}Aver, Erik, and Olive, Keith A., and Skillman, Evan D. \emph{The effects of He I {\ensuremath{\lambda}}10830 on helium abundance determinations.} \href{https://ui.adsabs.harvard.edu/abs/2015JCAP...07..011A}{JCAP \textbf{2015} (2015) pg. 011}

\bibitem{2014MNRAS.445..778I}Izotov, Y.I., Thuan, T.X., and Guseva, N.G. \emph{A new determination of the primordial He abundance using the He I {\ensuremath{\lambda}}10830 {\r{A}} emission line: cosmological implications} \href{https://ui.adsabs.harvard.edu/abs/2014MNRAS.445..778I}{MNRAS \textbf{445} (2014) pg. 778-793}

\bibitem{1982A&A...115..357S}Spite, F., and Spite, M. \emph{Abundance of lithium in unevolved stars and old disk stars: Interpretation and consequences.} \href{https://ui.adsabs.harvard.edu/abs/1982A&A...115..357S}{Astronomy and Astrophysics \textbf{115} (1982) pg. 357-366}

\bibitem{Ryan_2000}Ryan, Sean G.G., et al. \emph{Primordial Lithium and Big Bang Nucleosynthesis.} \href{https://doi.org/10.1086/312492}{ApJ \textbf{530} (2000) pg. L57}

\bibitem{Asplund_2006}Asplund, Martin, et al. \emph{Lithium Isotopic Abundances in Metal-poor Halo Stars} \href{https://doi.org/10.1086/503538}{ApJ \textbf{644} (2006) pg. 229}

\bibitem{2010A&A...522A..26S}Sbordone, L. et al. \emph{The metal-poor end of the Spite plateau. I. Stellar parameters, metallicities, and lithium abundances.} \href{https://ui.adsabs.harvard.edu/abs/2010A&A...522A..26S}{Astronomy and Astronomy \textbf{522} (2010) pg. A26}

\bibitem{1969ARA&A...7..553W}Wagoner, Robert V. \emph{Physics of Massive Objects.} \href{https://ui.adsabs.harvard.edu/abs/1969ARA&A...7..553W}{Annual Review of Astronomy and Astrophysics \textbf{7} (1969) pg. 553}

\bibitem{2017IJMPE..2641002C}Coc, Alain, and Vangioni, Elisabeth. \emph{Primordial nucleosynthesis.} \href{https://ui.adsabs.harvard.edu/abs/2017IJMPE..2641002C}{International Journal of Modern Physics E \textbf{26} (2017) pg. 1741002}

\bibitem{2020JCAP...03..010F}Fields, Brian D., et al. \emph{Big-Bang Nucleosynthesis after Planck.} \href{https://ui.adsabs.harvard.edu/abs/2020JCAP...03..010F}{JCAP \textbf{2020} (2020) pg. 010}

\bibitem{1937Natur.139..323D}Dirac, P.A.M. \emph{The Cosmological Constants.} \href{https://ui.adsabs.harvard.edu/abs/1937Natur.139..323D}{Nature \textbf{139} (1937) pg. 323}

\bibitem{1977ApJ...211..342C}Canuto, V., and Lodenquai, J. \emph{Dirac cosmology.} \href{https://ui.adsabs.harvard.edu/abs/1977ApJ...211..342C}{ApJ \textbf{211} (1977) pg. 342-356}

\bibitem{2023MNRAS.520.1447M}Maeder, Andre. \emph{MOND as a peculiar case of the SIV theory.} \href{https://ui.adsabs.harvard.edu/abs/2023MNRAS.520.1447M}{MNRAS \textbf{520} (2023) pg. 1447-1455}


\bibitem{2017ApJ...847...65M}Maeder, Andre. \emph{Scale-invariant Cosmology and CMB Temperatures as a Function of Redshifts.} \href{https://ui.adsabs.harvard.edu/abs/2017ApJ...847...65M}{ApJ \textbf{847} (2017) pg. 65}

\bibitem{2019arXiv190210115M}Maeder, Andre. \emph{Evolution of the early Universe in the scale invariant theory} \href{https://ui.adsabs.harvard.edu/abs/2019arXiv190210115M}{arXiv:1902.10115 \textbf{} (2019)}

\bibitem{1961PhRv..124..925B}Brans, C., and Dicke, R.H. \emph{Mach's Principle and a Relativistic Theory of Gravitation.} \href{https://ui.adsabs.harvard.edu/abs/1961PhRv..124..925B}{Physical Reviews \textbf{124} (1961) pg. 925-935}

\bibitem{1918SPAW.......465W}Weyl, Hermann. \emph{Gravitation und Elektrizit{\"a}t} \href{https://ui.adsabs.harvard.edu/abs/1918SPAW.......465W}{Sitzungsberichte der K{\"o}niglich Preussischen Akademie der Wissenschaften (1918) pg. 465-478}

\bibitem{1977PhRvL..39..429C}Canuto, V., Hsieh, S.H., and Adams, P.J. \emph{Scale-Covariant Theory of Gravitation and Astrophysical Applications.} \href{https://ui.adsabs.harvard.edu/abs/1977PhRvL..39..429C}{Physical Review Letters \textbf{39} (1977) pg. 429-432}

\bibitem{2012CQGra..29o5015R}Romero, C., Fonseca-Neto, J.B., and Pucheu, M.L \emph{General relativity and Weyl geometry.} \href{https://ui.adsabs.harvard.edu/abs/2012CQGra..29o5015R}{Classical and Quantum Gravity \textbf{29} (2012) pg. 155015}

\bibitem{2023Symm...15..709C}Cuzinatto, Rodrigo R., Gupta, Rajendra P., and Pompeia, Pedro J. \emph{Dynamical Analysis of the Covarying Coupling Constants in Scalar-Tensor Gravity} \href{https://ui.adsabs.harvard.edu/abs/2023Symm...15..709C}{Symmetry \textbf{15} (2023) pg. 709}

\bibitem{1956MNRAS.116..678G}Gilbert, C. \emph{The gravitational field of a star in the expanding universe.} \href{https://ui.adsabs.harvard.edu/abs/1956MNRAS.116..678G/abstract}{MNRAS \textbf{116} (1956) pg. 678}1956

\bibitem{1961Natur.192...57G} Gilbert, C. \emph{Dirac's Cosmology.} \href{https://ui.adsabs.harvard.edu/abs/1961Natur.192...57G/abstract}{Nature \textbf{192} (1961) pg. 57}

\bibitem{PhysRev.73.801}Teller, Edward. \emph{On the Change of Physical Constants} \href{https://link.aps.org/doi/10.1103/PhysRev.73.801}{Phys. Rev. \textbf{73} (1948) pg. 801-802}

\bibitem{PhysRevLett.36.833}Chin, Chao-wen, and Stothers, Richard. \emph{Limit on the Secular Change of the Gravitational Constant Based on Studies of Solar Evolution.} \href{https://link.aps.org/doi/10.1103/PhysRevLett.36.833}{Phys. Rev. Lett. \textbf{36} (1976)}

\bibitem{2014IJMPD..2342018S}Sahni, Varun, and Shtanov, Yuri. \emph{Can a variable gravitational constant resolve the faint young Sun paradox?} \href{https://ui.adsabs.harvard.edu/abs/2014IJMPD..2342018S}{International Journal of Modern Physics D \textbf{23} (2014) pg. 1442018-152}

\bibitem{1973Natur.241..519M}Morrison, L.V. \emph{Rotation of the Earth from AD 1663-1972 and the Constancy of G} \href{https://ui.adsabs.harvard.edu/abs/1973Natur.241..519M}{Nature \textbf{241} (1973) pg. 519-520}

\bibitem{PhysRevD.41.1034}Sisterna, P., and Vucetich, H. \emph{Time variation of fundamental constants: Bounds from geophysical and astronomical data.} \href{https://link.aps.org/doi/10.1103/PhysRevD.41.1034}{Phys. Rev. D \textbf{41} (1990) pg. 1034-1046}

\bibitem{Corsico_2013}Alejandro H. Córsico, Leandro G. Althaus, Enrique García-Berro, and Alejandra D. Romero. \emph{An independent constraint on the secular rate of variation of the gravitational constant from pulsating white dwarfs.} \href{https://doi.org/10.1088/1475-7516/2013/06/032}{Journal of Cosmology and Astroparticle Physics \textbf{2013} (2013) pg. 032}

\bibitem{1996A&A...312..345D}degl'Innocenti, S., Fiorentini, G., Raffelt, G.G., Ricci, B., and Weiss, A. \emph{Time-variation of Newton's constant and the age of globular clusters.} \href{https://ui.adsabs.harvard.edu/abs/1996A&A...312..345D}{Astronomy and Astrophysics \textbf{312} (1996) pg. 345-352}

\bibitem{10.1046/j.1365-8711.1999.02486.x}Benvenuto, O. G., Althaus, L. G. and Torres, Diego F. \emph{Evolution of white dwarfs as a probe of theories of gravitation: the case of Brans—Dicke} \href{https://doi.org/10.1046/j.1365-8711.1999.02486.x}{MNRAS \textbf{305} (1999) pg. 905-919}

\bibitem{1996PhRvL..77.1432T}Thorsett, S.E. \emph{The Gravitational Constant, the Chandrasekhar Limit, and Neutron Star Masses} \href{https://ui.adsabs.harvard.edu/abs/1996PhRvL..77.1432T}{PRL \textbf{77} (1996) pg. 1432-1435}

\bibitem{2004PhRvL..93z1101WH}Williams, J. G., Turyshev, S. G., and Boggs, D. H. \emph{Progress in Lunar Laser Ranging Tests of Relativistic Gravity.} \href{10.1103/PhysRevLett.93.261101}{Physical Review Letters \textbf{93} (2004) pg. 261101}

\bibitem{2015JCAP...10..029B}Bai, Yang, Salvado, Jordi, and Stefanek, Ben A. \emph{Cosmological constraints on the gravitational interactions of matter and dark matter.} \href{https://ui.adsabs.harvard.edu/abs/2015JCAP...10..029B}{JCAP \textbf{2015} (2015) pg. 029}

\bibitem{2017PTEP.2017d3E03O}Ooba, Junpei, Ichiki, Kiyotomo, Chiba, Takeshi, and Sugiyama, Naoshi. \emph{Cosmological constraints on scalar-tensor gravity and the variation of the gravitational constant.} \href{https://ui.adsabs.harvard.edu/abs/2017PTEP.2017d3E03O}{Progress of Theoretical and Experimental Physics \textbf{2017} (2017) pg. 043E03}

\bibitem{2004PhRvL..92q1301C}Copi, Craig J., Davis, Adam N., and Krauss, Lawrence M. \emph{New Nucleosynthesis Constraint on the Variation of G.} \href{https://ui.adsabs.harvard.edu/abs/2004PhRvL..92q1301C}{PRL \textbf{92} (2004) pg. 171301}

\bibitem{2020EPJC...80..148A}Alvey, J., Sabti, N., Escudero, M., and Fairbairn, M. \emph{Improved BBN constraints on the variation of the gravitational constant.} \href{https://ui.adsabs.harvard.edu/abs/2020EPJC...80..148A}{European Physical Journal C \textbf{80} (2020) pg. 148}

\bibitem{2019ApJ...887L...1B}Bellinger, E. P., and Christensen-Dalsgaard, J. \emph{Asteroseismic Constraints on the Cosmic-time Variation of the Gravitational Constant from an Ancient Main-sequence Star.} \href{https://ui.adsabs.harvard.edu/abs/2019ApJ...887L...1B}{ApJL \textbf{887} (2019) pg. L1}

\bibitem{2004PhRvL..93z1101W}Williams, James G., Turyshev, Slava G., and Boggs, Dale H. \emph{Progress in Lunar Laser Ranging Tests of Relativistic Gravity.} \href{https://ui.adsabs.harvard.edu/abs/2004PhRvL..93z1101W}{PRL \textbf{93} (2004) pg. 261101}

\bibitem{2018CQGra..35c5015H}Hofmann, F., and M{\"u}ller, J. \emph{Relativistic tests with lunar laser ranging.} \href{https://ui.adsabs.harvard.edu/abs/2018CQGra..35c5015H}{Classical and Quantum Gravity \textbf{35} (2018) pg. 035015}

\bibitem{2013MNRAS.432.3431P}Pitjeva, E.V., and Pitjev, N.P. \emph{Relativistic effects and dark matter in the Solar system from observations of planets and spacecraft.} \href{https://ui.adsabs.harvard.edu/abs/2013MNRAS.432.3431P}{MNRAS \textbf{432} (2013) pg. 3431-3437}

\bibitem{2015CeMDA.123..325F}Fienga, A., Laskar, J., Exertier, P., Manche, H., and Gastineau, M. \emph{Numerical estimation of the sensitivity of INPOP planetary ephemerides to general relativity parameters.} \href{https://ui.adsabs.harvard.edu/abs/2015CeMDA.123..325F}{Celestial Mechanics and Dynamical Astronomy \textbf{123} (2015) pg. 325-349}

\bibitem{2018NatCo...9..289G}Genova, Antonio, Mazarico, Erwan, Goossens, Sander, Lemoine, Frank G., Neumann, Gregory A., Smith, David E., and Zuber, Maria T. \emph{Solar system expansion and strong equivalence principle as seen by the NASA MESSENGER mission.} \href{https://ui.adsabs.harvard.edu/abs/2018NatCo...9..289G}{Nature Communications \textbf{9} (2018) pg. 289}

\bibitem{1988PhRvL..61.1151D}Damour, Thibault, and Gibbons, Gary W., and Taylor, Joseph H. \emph{Limits on the variability of G using binary-pulsar data.} \href{https://ui.adsabs.harvard.edu/abs/1988PhRvL..61.1151D}{PRL \textbf{61} (1988) pg. 1151-1154}

\bibitem{1994ApJ...428..713K}Kaspi, V.M., Taylor, J.H., and Ryba, M.~F. \emph{High-Precision Timing of Millisecond Pulsars. III. Long-Term Monitoring of PSRs B1855+09 and B1937+21.} \href{https://ui.adsabs.harvard.edu/abs/1994ApJ...428..713K}{ApJ \textbf{428} (1994) pg. 713}

\bibitem{2019MNRAS.482.3249Z}Zhu, W.W., et al. \emph{Tests of gravitational symmetries with pulsar binary J1713+0747.} \href{https://ui.adsabs.harvard.edu/abs/2019MNRAS.482.3249Z}{MNRAS \textbf{482} (2019) pg. 3249-3260}

\bibitem{2001PhRvD..65b3506G}Gazta{\~n}aga, E., Garc{\'\i}a-Berro, E., Isern, J., Bravo, E., and Dom{\'\i}nguez, I. \emph{Bounds on the possible evolution of the gravitational constant from cosmological type-Ia supernovae.} \href{https://ui.adsabs.harvard.edu/abs/2001PhRvD..65b3506G}{PRD \textbf{65} (2001) pg. 023506}

\bibitem{2018PhRvD..97h3505W}Wright, Bill S., and Li, Baojiu. \emph{Type Ia supernovae, standardizable candles, and gravity} \href{https://ui.adsabs.harvard.edu/abs/2018PhRvD..97h3505W}{{RD} \textbf{97} (2018) pg. 083505}

\bibitem{1907AnP...328..197E}Einstein, A. \emph{{\"U}ber die M{\"o}glichkeit einer neuen Pr{\"u}fung des Relativit{\"a}tsprinzips} \href{https://ui.adsabs.harvard.edu/abs/1907AnP...328..197E}{nnalen der Physik \textbf{328} (1907) pg. 197-198}

\bibitem{1957RvMP...29..363D}Dicke, R.H. \emph{Gravitation without a Principle of Equivalence.} \href{https://ui.adsabs.harvard.edu/abs/1957RvMP...29..363D}{Reviews of Modern Physics \textbf{29} (1957) pg. 363-376}

\bibitem{1993gr.qc....12017M}Moffat, J.W. \emph{Predictability in Quantum Gravity and Black Hole Evaporation} \href{https://ui.adsabs.harvard.edu/abs/1993gr.qc....12017M}{arXiv:gr-qc/9312017 (1993)}

\bibitem{1993gr.qc.....6003M}Moffat, J.W. \emph{Consistency of the Nonsymmetric Gravitational Theory} \href{https://ui.adsabs.harvard.edu/abs/1993gr.qc.....6003M}{arXiv:gr-qc/9306003 (1993)}

\bibitem{1999PhRvD..59d3516A}Albrecht, Andreas, and Magueijo, Joao. \emph{Time-varying speed of light as a solution to cosmological puzzles.} \href{https://ui.adsabs.harvard.edu/abs/1999PhRvD..59d3516A}{PRD \textbf{59} (1999) pg. 043516}

\bibitem{1999PhRvD..59d3515B}Barrow, John D. \emph{Cosmologies with varying light speed.} \href{https://ui.adsabs.harvard.edu/abs/1999PhRvD..59d3515B}{PRD \textbf{59} (1999) pg. 043515}

\bibitem{1999PhLB..459..468A}Avelino, P.P. and Martins, C.J.A.P. \emph{Does a varying speed of light solve the cosmological problems?} \href{https://ui.adsabs.harvard.edu/abs/1999PhLB..459..468A}{Physics Letters B \textbf{459} (1999) pg. 468-472}

\bibitem{2000PhRvD..62l3508A}Avelino, P.P., Martins, C.J.A.P., Rocha, G., and Viana, P. \emph{Looking for a varying {\ensuremath{\alpha}} in the cosmic microwave background.} \href{https://ui.adsabs.harvard.edu/abs/2000PhRvD..62l3508A}{PRD \textbf{62} (2000) pg. 123508}

\bibitem{2016EPJC...76..130M}Moffat, J.W. \emph{Variable speed of light cosmology, primordial fluctuations, and gravitational waves.} \href{https://ui.adsabs.harvard.edu/abs/2016EPJC...76..130M}{European Physical Journal C \textbf{76} (2016) pg. 130}

\bibitem{2024ApJ...964...55G}Gupta, Rajendra P. \emph{Testing CCC+TL Cosmology with Observed Baryon Acoustic Oscillation Features} \href{https://ui.adsabs.harvard.edu/abs/2024ApJ...964...55G}{ApJ \textbf{964} (2024) pg. 55}

\bibitem{2024Univ...10..266G}Gupta, Rajendra P. \emph{On Dark Matter and Dark Energy in CCC+TL Cosmology} \href{https://ui.adsabs.harvard.edu/abs/2024Univ...10..266G}{Universe \textbf{10} (2024) pg. 266}

\bibitem{2025Galax..13..108G}Gupta, Rajendra P.\emph{Testing CCC+TL Cosmology with Galaxy Rotation Curves} \href{https://ui.adsabs.harvard.edu/abs/2025Galax..13..108G}{Galaxies \textbf{13} (2025) pg. 108}

\bibitem{2025Galax..13..115G}Gupta, Rajendra P. \emph{Evolution of Size, Mass, and Density of Galaxies Since Cosmic Dawn} \href{https://ui.adsabs.harvard.edu/abs/2025Galax..13..115G}{Galaxies \textbf{13} (2025) pg. 115}

\bibitem{1921SPAW.......966K}Kaluza, Theodor. \emph{Zum Unit{\"a}tsproblem der Physik} \href{https://ui.adsabs.harvard.edu/abs/1921SPAW.......966K}{Sitzungsberichte der K{\"o}niglich Preussischen Akademie der Wissenschaften (1921) pg. 966-972}

\bibitem{1967JETPL...5...24S}Sakharov, A.D. \emph{Violation of CP Invariance, C Asymmetry, and Baryon Asymmetry of the Universe.} \href{https://ui.adsabs.harvard.edu/abs/1967JETPL...5...24S}{Soviet Journal of Experimental and Theoretical Physics Letters \textbf{5} (1967) pg. 24}

\bibitem{1997PhR...283..303O}Overduin, J.M. and Wesson, P.S. \emph{Kaluza-Klein gravity.} \href{https://ui.adsabs.harvard.edu/abs/1997PhR...283..303O}{Phys. Rep. \textbf{283} (1997) pg. 303-378}

\bibitem{2010grav.book.....P}Padmanabhan, T. \emph{Gravitation: Foundations and Frontiers} \href{https://ui.adsabs.harvard.edu/abs/2010grav.book.....P}{(2010)}

\bibitem{1994GReGr..26.1171D}Damour, T., and Polyakov, A.M. \emph{String theory and gravity.} \href{https://ui.adsabs.harvard.edu/abs/1994GReGr..26.1171D}{General Relativity and Gravitation \textbf{26} (1994) pg. 1171-1176}

\bibitem{2003AnHP....4S.347U}Uzan, Jean-Philippe. \emph{Tests of Gravity on Astrophysical Scales and Variation of the Constants} \href{https://ui.adsabs.harvard.edu/abs/2003AnHP....4S.347U}{Annales Henri Poincar{\'e} \textbf{4} (2003) pg. 347-369}

\bibitem{2017RPPh...80l6902M}Martins, C.J.A.P. \emph{The status of varying constants: a review of the physics, searches, and implications.} \href{https://ui.adsabs.harvard.edu/abs/2017RPPh...80l6902M}{Reports on Progress in Physics \textbf{12} (2017) pg. 126902}

\bibitem{2023MNRAS.526.3987C}Cuzinatto, R.R., de Melo, C.A.M., and Neves, Juliano C.S. \emph{Shadows of black holes at cosmological distances in the co-varying physical couplings framework.} \href{https://ui.adsabs.harvard.edu/abs/2023MNRAS.526.3987C}{MNRAS \textbf{526} (2023) pg. 3987-3993}

\bibitem{2023AAS...24232604G}Gupta, Rajendra P. \emph{JWST: 'Impossible Early Galaxy' Problem and {\ensuremath{\Lambda}}CDM Cosmology.} \href{https://ui.adsabs.harvard.edu/abs/2023AAS...24232604G}{American Astronomical Society Meeting Abstracts \#242 \textbf{242} (2023) pg. 326.04}

\bibitem{2026Symm...18..300G}Gupta, Rajendra P.  \emph{The Origin of Dark Matter and Dark Energy: Covarying Coupling Constants?} \href{https://ui.adsabs.harvard.edu/abs/2026Symm...18..300G}{Symmetry \textbf{18} (2026) pg. 300}

\bibitem{2022MPLA...3750155G}Gupta, Rajendra P. \emph{Varying coupling constants and their interdependence.} \href{https://ui.adsabs.harvard.edu/abs/2022MPLA...3750155G}{Modern Physics Letters A \textbf{37} (2022) pg. 2250155}

\bibitem{2018ApJ...859..101S}Scolnic, D.M. et al. \emph{The Complete Light-curve Sample of Spectroscopically Confirmed SNe Ia from Pan-STARRS1 and Cosmological Constraints from the Combined Pantheon Sample.} \href{https://ui.adsabs.harvard.edu/abs/2018ApJ...859..101S}{ApJ \textbf{859} (2018) pg. 101}

\bibitem{2022ApJ...938..110B}Brout, Dillon. et al. \emph{The Pantheon+ Analysis: Cosmological Constraints} \href{https://ui.adsabs.harvard.edu/abs/2022ApJ...938..110B}{ApJ \textbf{938} (2022) pg. 110}

\bibitem{1989RvMP...61...25B}Bernstein, Jeremy, Brown, Lowell S., and Feinberg, Gerald. \emph{Reviews of Modern Physics} \href{https://ui.adsabs.harvard.edu/abs/1989RvMP...61...25B}{Reviews of Modern Physics \textbf{61} (1989) pg. 25-39}

\bibitem{1990eaun.book.....K}Kolb, Edward W., and Turner, Michael S. \emph{The early universe} \href{https://ui.adsabs.harvard.edu/abs/1990eaun.book.....K}{\textbf{69} (1990)}

\bibitem{2014arXiv1411.3687W}Wietfeldt, F.E. \emph{The Neutron Lifetime} \href{https://ui.adsabs.harvard.edu/abs/2014arXiv1411.3687W}{arXiv:1411.3687 (2014)}

\end{thebibliography}

\end{document}